\documentclass[preprint,sort&compress,12pt]{elsarticle}

\usepackage{amssymb}
\usepackage{natbib}
\usepackage{amsmath,amsthm,bm,mathrsfs}

 \newtheoremstyle{theorem}{6pt}{6pt}{\rm}{}{\sffamily}{ }{ }{}
 \theoremstyle{theorem}

 \newtheoremstyle{algorithm}{6pt}{6pt}{\rm}{}{\sffamily}{ }{ }{}
 \theoremstyle{algorithm}

 \newtheoremstyle{lemma}{6pt}{6pt}{\rm}{}{\sffamily}{ }{ }{}
 \theoremstyle{lemma}

\newtheoremstyle{case}{6pt}{6pt}{\rm}{}{\sffamily}{. }{ }{}
 \theoremstyle{case}

 \newtheoremstyle{statement}{6pt}{6pt}{\rm}{}{\sffamily}{ }{ }{}
\theoremstyle{statement}

 \newtheoremstyle{corollary}{6pt}{6pt}{\rm}{}{\sffamily}{ }{ }{}
 \theoremstyle{corollary}

  \newtheoremstyle{definition}{6pt}{6pt}{\rm}{}{\sffamily}{ }{ }{}
 \theoremstyle{definition}
 \newtheorem{definition}{\sc Definition}[section]

\newtheoremstyle{example}{6pt}{6pt}{\rm}{}{\sffamily}{ }{ }{}
\theoremstyle{example}

\newtheoremstyle{remark}{6pt}{6pt}{\rm}{}{\sffamily}{ }{ }{}
\theoremstyle{remark}

\newtheoremstyle{approximation}{6pt}{6pt}{\rm}{}{\sffamily}{ }{ }{}
\theoremstyle{approximation}

\newtheoremstyle{scheme}{6pt}{6pt}{\rm}{}{\sffamily}{ }{ }{}
\theoremstyle{scheme}

\newtheoremstyle{Algorithm}{6pt}{6pt}{\rm}{}{\sffamily}{ }{ }{}
\theoremstyle{Algorithm}

\newtheoremstyle{Assumption}{6pt}{6pt}{\rm}{}{\sffamily}{ }{ }{}
\theoremstyle{Assumption}

\newtheoremstyle{proposition}{6pt}{6pt}{\rm}{}{\sffamily}{ }{ }{}
\theoremstyle{proposition}

\newtheoremstyle{hypo}{6pt}{6pt}{\rm}{}{\sffamily}{ }{ }{}
 \theoremstyle{hypo}

  \newtheoremstyle{Step}{6pt}{6pt}{\rm}{}{}{ }{ }{}
 \theoremstyle{Step}

\journal{Physica A}

\begin{document}

\begin{frontmatter}

\title{Information theoretic approach for accounting classification}
\author{E.M.S. Ribeiro} 
\ead{esaidel@usp.br}
\author{G.A. Prataviera}
\ead{prataviera@usp.br}
\address{Departamento de Administra\c{c}\~ao, FEARP, Universidade de S\~{a}o Paulo, 14040-905, Ribeir\~{a}o Preto, SP, Brazil}

\begin{abstract}
In this paper we consider an information theoretic approach 
for the accounting classification process. We propose a matrix formalism and an 
algorithm for calculations of information theoretic measures associated to accounting classification.
The formalism may be useful for further generalizations and computer-based implementation. Information 
theoretic measures, mutual information and symmetric uncertainty, were evaluated for daily transactions recorded 
in the chart of accounts of a small company during two years. Variation in the information measures due the aggregation of data in the process of accounting classification is observed. In particular, the symmetric uncertainty seems to be a useful parameter for comparing companies over time or in different sectors or different accounting choices and standards.
\end{abstract}


\end{frontmatter}



%this files contains Theorem styles based in IMA JOURNALS
%\input standard.tex

%%%%%%%%%%%%%%section A%%%%%%%%%
\section{Introduction}
%%%%%%%%%%%%%%%%%%%%%%%%%%%%%

Information Theory \cite{shannon1} provides useful and unifying concepts that have been applied in many fields, 
including Physics, Engineering, Computer science, Statistics and Data Analysis, Linguistics, Marketing, Economics, and Complex Systems research in general \cite{jaynes,inter9,inter6,inter7,inter3,inter8,mackay,coverthomas,kulback,inter2,inter10,inter5,brockett2,brockett,golan,inter1,inter4,haken}. Here, we are including Accounting as a potential field for interdisciplinary research where concepts and methods from information theory may have interesting applications. In fact, Accounting is considered an information science used to collect, classify, and manipulate financial data for organizations and individuals~\cite{demski}. Nevertheless, an approach based on information theory was reported only in some academic research studies from 1960’s \cite{bedbala,bostwick,leebedford,lev, theil}. Specifically, in ref~\cite{bedbala} the authors proposed the viewing of Accounting as a 
communication process; Lee and Bedford~\cite{leebedford} proposed a communication channel 
model to describe accounting classification. However, as far as we know, neither applications using real data nor further 
research in the field was performed.

Accounting classification is a relevant subject for international 
accounting harmonization studies \cite{darcy,nobes,nobes2}. In~\cite{darcy}, 
the author observed a high level of complexity of an accounting system due 
to the conceptual and methodological pluralism found in accounting 
classification attempts. By considering fifteen national systems a cluster analysis was performed and 
a nonmetric multidimensional scaling technique was applied, obtaining a two-dimensional map revealing similarities 
(dissimilarities) between the systems. In ~\cite{nobes} and ~\cite{nobes2} the author investigated 
international differences in the way that countries and companies have 
responded to the International Financial Reporting Standards (IFRS). They conclude 
that some countries have entirely abandoned national accounting rules in favor 
of IFRS. It is also observed that different national systems of IFRS practices are emerging, 
and will only be classified further.. Thus, a quantitative measure of information 
may provide a more objective way to compare different accounting systems.

As pointed out by Demski~\cite{demski}, there is absence of modern information science 
in the Accounting curriculum. In the research domain, there are several examples 
involving Accounting that can be related to probability and allowing for an information 
theoretic perspective \cite{christensen, willet1}. Actually, 
the value of information \cite{epstein} and the information content of inside 
traders before and after the Sarbanes-Oxley Act of 2002 (SOX) \cite{brochet} have 
been previously considered. However, no investigation on information 
based on information theory was done. Hence, if compared with applications in other areas, information theory was not 
enough explored in Accounting. Besides, the technological developments including the storage and access of data, computer time processing, and the evolutions of information systems may allow the implementation and inclusion 
of results that previously were not possible. Therefore, attempting to develop new tools to support further research, it seems appropriate to review and further explore earlier studies.

The aim of this work is to further explore the information theoretic approach for accounting classification. A matrix formalism and an algorithm for calculations of information theoretic measures are introduced. Although the objects were correctly defined in Ref.~\cite{leebedford}, a matrix formalism was avoided. Our formalism may be useful for further generalizations and computer-based implementations. The formalism is applied to evaluate the information theoretic measures in classifying the transactions of a small company. The mutual information and the symmetric uncertainty are obtained for each level of classification in the chart of accounts, allowing us to observe their variation due to aggregation of data in the process of accounting classification. To the best of our knowledge, this is the first calculation of information theoretic measures for an accounting classification process using an empirical data set. Moreover, we indicate that the symmetric uncertainty may be a useful parameter for comparing companies over time or in different sectors or different accounting choices and standards.

The article is organized as follows: in section \ref{sec.concepts} some basic 
concepts of information theory are presented. In section \ref{sec.theory} the work of 
Lee and Bedford connecting the processes of accounting classification and the theory 
of information is revisited. The matrix formalism and an algorithm for calculations 
are described in section \ref{sec.algorith}. In section \ref{sec.application}, information theoretic measures are evaluated considering the events registered in a five level chart of accounts of a small company. In section \ref{sec.conclusion} some concluding remarks are presented. Finally, the source code for the R software environment \cite{softR} is included as an Appendix.
%%%%%%%%%%%%%%%%%%%%%%%%%%%%%%%%%%%%%%%%%%%%%%%%%%%%%%%%%%%%%%%%%%%%%%%%%%%%%%%%%%%%%%%%%%%
\section{Shannon entropy and the measure of information}{\label{sec.concepts}}
In this section some basic definitions of information theory are presented. A complete and more detailed treatment
may be found elsewhere \cite{coverthomas,mackay}. 

Given an event $A$ occurring with probability $P(A)$, it is possible to associate a number, 
$-\log_2 P(A)$, to quantify the information associated with the occurrence of $A$. 
This definition agrees with the intuitive idea that the information content of 
independent events is the sum of the information of each event. In order to 
quantify the information content of a set of events, Shannon introduced the concept 
of average amount of information or entropy~\cite{shannon1}.

\begin{definition}
Given the set of events $X=\left\{x_1, x_2, \ldots, x_n \right\}$ with probabilities 
$\left\{P(x_1),\right.$ $\left. P(x_2), \ldots,P(x_n)\right\}$, the entropy $H(X)$ 
associated to $X$ is defined as the mean information of $X$:
\begin{equation}
H(X)=-\sum_{i=1}^{n} P(x_i) \log_{2} P(x_i). 
\end{equation}
\end{definition}
Entropy may be interpreted as a measure of the uncertainty associated to a set of 
random events. The unit of information using the logarithm function to base two is 
called bit. Since $P(x_i)$ may be zero, $H(X)$ could be indeterminate in the above 
definition so, when $P(x_i)=0$, the value zero is assigned to $P(x_i) \log_{2} P(x_i)$. 
For two or more sets of events described by a joint probability distribution, the 
joint entropy may be defined as follows.
\begin{definition}
Given two sets $(X,Y)$ of random events, with the joint probability 
distribution $P(X,Y)$, the joint entropy between $X$ and $Y$ is defined as
\begin{equation}
H(X,Y)=-\sum_{i=1}^{n}\sum_{j=1}^{m} P(x_i,y_j) \log_{2} P(x_i,y_j),
\end{equation}  
where $n$ and $m$ are the total number of distinct events in the $X$ and $Y$ sets, respectively.
\end{definition}
For independent events the joint probability distribution factorizes and the joint entropy 
becomes the sum of the entropy of each set of events, i.e., 
$H(X,Y)=H(X)+H(Y)$. Moreover, for joint distributions, if the information 
about one variable is conditioned to the information about another variable, 
it is useful to define the conditional entropy.
\begin{definition}
Given two sets $(X,Y)$ of random events, the conditional entropy $H(X/Y)$ 
is defined as
\begin{equation}
H(X/Y)=-\sum_{i=1}^{n}\sum_{j=1}^{m} P(x_i,y_j) \log_{2} P(x_i/y_j), 
\end{equation} 
where $P(x_i/y_j)$ is the conditional probability of a random variable $X$ 
which assumes the value $x_i$, given that another random variable $Y$ has taken a 
value $y_j$.
\end{definition}
The conditioned entropy will be useful in the accounting classification 
process since it involves the registration of an economic event in one account 
given that it comes from another account \cite{leebedford}, i.e., a double entry system. The 
joint entropy may be expressed in terms of the conditional entropy as 
$H(X,Y)= H(X/Y)+H(Y)$, and since $H(X/Y)\leq H(X)$, it follows that 
$H(X,Y)\leq H(X)+H(Y)$.
 
We conclude this section with the concept of mutual information 
$I(X,Y)$, which is the amount of information that one random variable 
contains about another random variable.
\begin{definition}
Given two sets $(X,Y)$ of random events, the mutual information $I(X/Y)$ 
between $X$ and $Y$ is defined as
\begin{equation}
I(X,Y)= H(X)-H(X/Y)=H(X)+H(Y)-H(X,Y). 
\end{equation}
\end{definition}
Mutual information is the reduction in the uncertainty of one random variable 
due to the knowledge of the other \cite{coverthomas}. By considering the 
joint probability distribution the mutual information may be written as 
\begin{equation}
I(X,Y)=-\sum_{i=1}^{n}\sum_{j=1}^{m} P(x_i,y_j) \log_{2} \frac{P(x_i,y_j)}{P(x_i)P(y_j)}.
\end{equation}
For independent events, the joint probability distribution factorizes as 
$P(X,Y)$ $ = $ $P(X)P(Y)$, and $I(X,Y)=0$. Thus, mutual information may be used as a 
measure of the degree of association between random events. It is also useful to consider a normalized 
form of Mutual Information \cite{witten}, the symmetric uncertainty, which 
is basically a measure of correlation defined as
\begin{equation}
U(X,Y)=2\frac{I(X,Y)}{H(X)+H(Y)}.
\end{equation}
Symmetric uncertainty lies between 0 and 1, and it is the information
shared between $X$ and $Y$ relatively at all information contained in both $X$
and $Y$.

\section{Accounting classification and information theory}
\begin{figure}[ht!]
\centering
\includegraphics[width=1.0\textwidth]{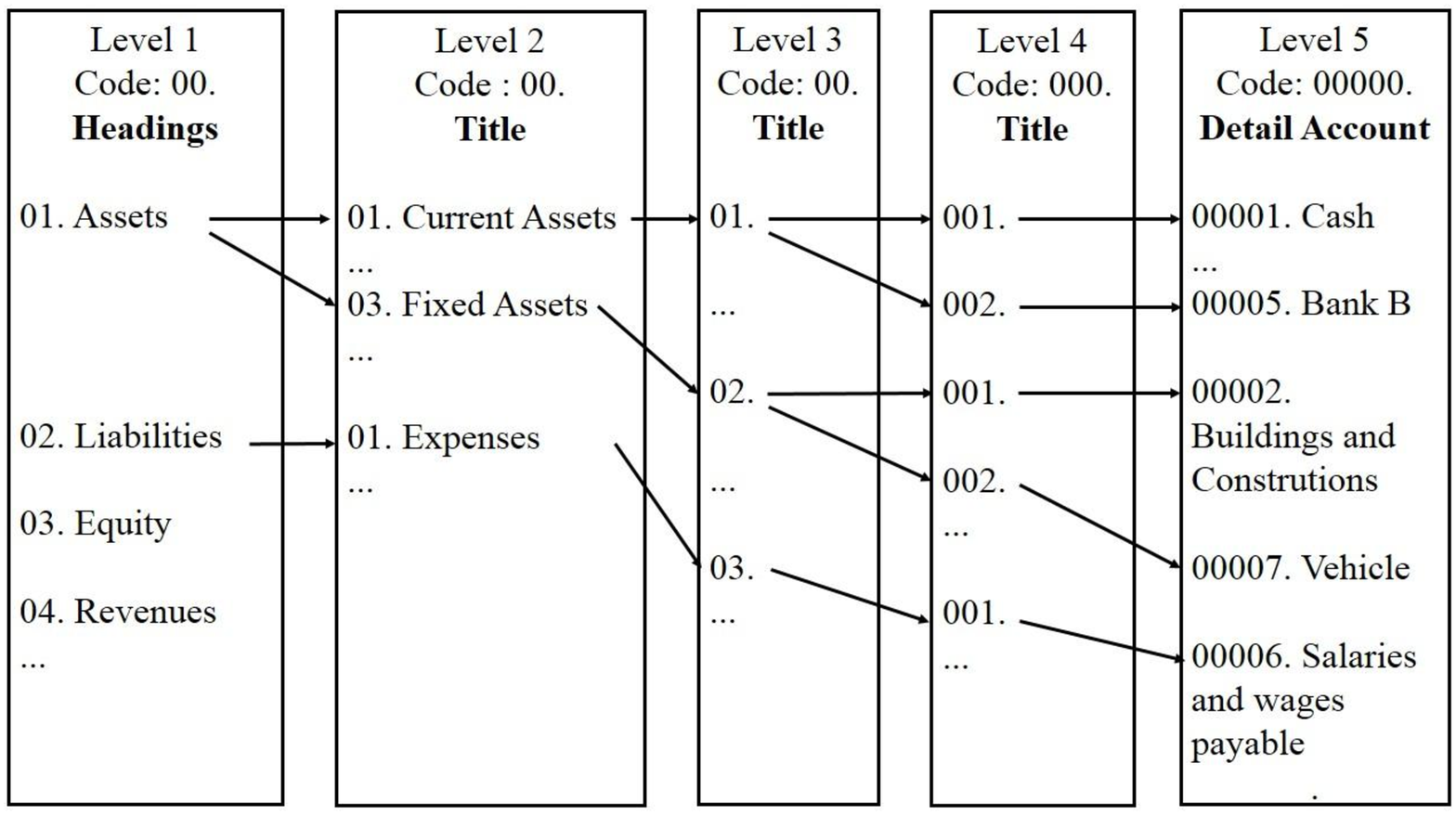}
\caption{Illustration of a typical chart of accounts with five levels. Here, economic events are encoded by a sequence of numbers with five pieces separated by dots.}
\label{fig.coa}
\end{figure}

{\label{sec.theory}}
The mathematical model for the accounting classification process and its connection 
with information theory were first introduced by Lee and Bedford~\cite{leebedford}. The overall process can 
be formalized by means of matrix algebra as 
will be shown in this paper. The first step of an accounting process in a firm is the registration of 
economic events.
\begin{definition}\rm
An elementary economic event is defined as any activity that an accountant 
records. The economic events of a firm are represented by the set 
$X=\left\{x_1, x_2, \ldots, x_R \right\}$ of elementary economic events.
\end{definition}
The accounting classification involves the registration of economic events 
as a debit or a credit. Then, the events may have several possibilities of 
classification.
\begin{definition}\rm  
The set $Y=\left\{y_1, y_2, \ldots, y_S \right\}$ represents the  possibilities 
of classification as a debit in one account $a_i$ and a credit in another 
account $a_j$. The maximum number of classifications in $N_a$ accounts is given 
by $ S = N_a(N_a-1)$ , which is the number of permutations of $N_a$ accounts 
taken two at a time.
\end{definition}
The function of accountant is to designate, among $N_a$ accounts, which one represents 
the structure of the financial state, with at least 
two accounts involved in each transaction \cite{leebedford}. With the 
definitions of the economic events set of a firm and the classification set it is now possible to define the accounting classification process.
\begin{definition}\rm 
The accounting classification process is a map $f:X\rightarrow Y$ relating 
each economic event $x_i$ to an element $y_j$.
\label{account.class} 
\end{definition}
Theory of information is based on the concept of information content of a set 
of events described by a probability distribution. In order to quantify the 
information content of an account transaction one needs to introduce the 
probability associated to the classification process. In Ref.\cite{leebedford} the probabilities were introduced as subjective numbers depending on the accountant decision. We remove the subjectivity by looking directly to the transaction frequencies registered by the company. These distributions of frequencies should reflect the effects of accounting standards (or even the accountant subjectivity). Then, the probability of an economic event is identified with its relative frequency, i. e.,
\begin{definition}\rm
The probability associate to an account transaction of an event $x_i$ is given by $P(x_i)= n_i / R$,
where $n_i$ is number of occurrences of the event  $x_i$ in the total number of events $R$.
\end{definition}
Indeed, such probabilities can be obtained from some information system. The set of economic events is organized in the so-called chart of accounts~\cite{acc.coach}, a listing of the account names $a_i$ that a company has identified and made available for recording transactions in its general ledger. A company has the flexibility to tailor its chart of accounts to best suit its needs. A specific economic event in the chart of accounts is coded as a sequence of numbers. Figure~\ref{fig.coa} illustrates a typical structure of a chart of accounts with five levels encoded in a string of numbers with five pieces separated by dots. At the last level (Level 5) each economic event receives a specific code. For example, from Figure~\ref{fig.coa} the event ``Salaries and Wages Payable" is encoded as 02.01.03.001.00006. At the intermediate levels the aggregation increases until reaching the first level (Level 1), which is the most aggregated one containing the events grouped in the main accounts. The classification process can be analyzed at each level, by 
introducing the conditional probability that a given economic event $x_i$ is classified 
as $y_j$, namely $P(y_j/x_i)$.

The attribution of probabilities to the classification event of an account transaction permits us to associate quantities such as entropy and mutual information to characterize the process of accounting classification at 
different levels. Then, at least formally, the accounting classification process can be analyzed as a process of gain and loss of information.
 
A communication channel is a system in which the output depends probabilistically 
on its input. It is characterized by the probability transitions $P(Y/X)$ 
that determines the conditional distribution of the output given the input. To summarize, 
the following association of accounting classification and information theory is 
possible: $H(X)$ is the entropy associated to the set of economic events $X$, $H(Y)$ 
is the entropy associated to the classification of the economic events in the set 
of classifications $Y$, $H(X, Y)$ is the mutual entropy between the economic events 
and its classification as a debit or credit in the set $Y$, and $H(X/Y)$ is the entropy 
associated to the economic events $X$ given their classification 
$Y$. Finally, $I(X,Y)$ is the average information about $X$ conveyed through the 
channel of classification $Y$.
\section{Algorithm for information analysis}
\begin{figure}[ht!]
\centering
\includegraphics[width=0.50\textwidth]{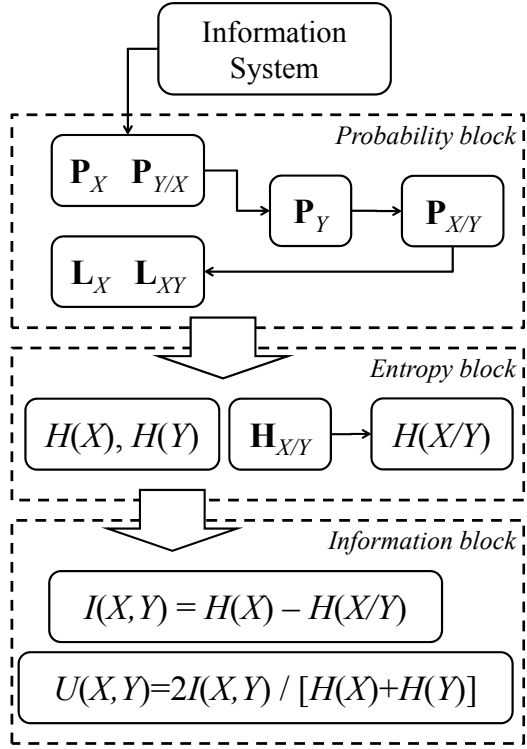}
\caption{Flowchart to obtain information measures associated to accounting classification process.}
\label{fig.chart}
\end{figure}

{\label{sec.algorith}}
The accounting process of classification is, in this way, an information system in 
which the data from economic events are aggregated into specific 
accounts. During the accounting classification, the transmission of information is 
realized through a communication channel \cite{bedbala}. In order to obtain the information 
measures associated to the process of accounting classification we propose a matrix 
formalism and an algorithm for calculations. The steps to obtain the amount of 
information transmitted in this channel can be structured in the flowchart of 
Figure~\ref{fig.chart}. The goal in applying this algorithm is to obtain the average information 
about $X$ conveyed through the channel by $Y$, namely $I(X,Y)$. We have assumed that ${\bf P}_X$, the probabilities for the occurrence of economic events $X$, as well as ${\bf P}_{Y/X}$, the probability that the 
events $X$ are classified as $Y$, is obtained from their frequencies of occurrence and supplied by some information system. All others quantities are obtained from these inputs. In the following the mathematical expressions and the matrix algebra to obtain the quantities are presented.

\subsection{The probability block}
This first algorithm block consists of, from the input probability matrices, 
determining additional probability matrices, suitable to be used in the 
determination of entropy functions (see Fig.~\ref{fig.chart}). The first 
input is the ${\bf P}_X$ matrix, whose entries are given by the probability 
of occurrence of each distinct economic event $x_i$, namely $P(x_i)$. 
In this way, considering the maximum number of economic events - $R$, 
${\bf P}_X$ can be written as an $1\times R$ matrix 
\begin{equation}
{\bf P}_{X} =\left[ 
\begin{array}{cccc} 
P(x_1) & P(x_2) & \cdots & P(x_R)%
\end{array}%
\right] .  
\end{equation}
Another input in this algorithm is ${\bf P}_{Y/X}$, a $R\times S$ matrix,
with $S$ the maximum number of classifications, and entries given by the 
probability that an event $x_i$ is classified as $y_j$, 
i.e., 
\begin{equation}
{\bf P}_{Y/X} =\left[
\begin{array}{cccc}
P(y_1/x_1) & P(y_2/x_1) & \cdots & P(y_S/x_1)  \\ 
P(y_1/x_2) & P(y_2/x_2) & \cdots & P(y_S/x_2)  \\ 
\vdots & \vdots & \ddots & \vdots  \\
P(y_1/x_R) & P(y_2/x_R) & \cdots & P(y_S/x_R)  \\
\end{array}%
\right] .  
\end{equation}	
These two probabilities matrices, ${\bf P}_X$ and ${\bf P}_{Y/X}$, can 
be obtained from the information system, but more generally their 
elements can be viewed as parameters for theoretical studies on 
classification. In fact, the elements are related to the
economic event frequencies, and the adopted accounting 
classification procedure. 
 
Following the flowchart in Fig.~\ref{fig.chart}, the $1\times S$ 
probability matrix, ${\bf P}_{Y}=\left[P(y_1)\,\,\,P(y_2)\,\cdots P(y_S)
\,\,\right]$ can be 
calculated from the input by a matrix multiplication defined by
\begin{equation}
{\bf P}_{Y}={\bf P}_{X}{\bf P}_{Y/X}. 
\end{equation}
The last probability matrix to be determined in this block is the 
conditional probability matrix ${\bf P}_{X/Y}$, a matrix defined 
by the product
\begin{equation}{\label{pxgydef}}
{\bf P}_{X/Y}={\bf P}_{X_{D}}{\bf P}_{Y/X}{\bf P}_{Y^{-1}_D}. 
\end{equation}
In Eq. (\ref{pxgydef}), ${\bf P}_{X_D}$ and ${\bf P}_{Y^{-1}_D}$ are $R\times R$ 
and $S\times S$ diagonal matrices with entries 
 $[{\bf P}_{X_{D}}]_{i,j}=P(x_i)\delta_{i,j}$, ($i,j=1,2,...,R$)
and $[{\bf P}_{Y^{-1}_D}]_{i,j}=\delta_{i,j}/P(y_i)$, ($i,j=1,2,...,S$), 
respectively, and where $\delta_{i,j}$ is the usual Kronecker function, 
which is $0$ unless $i=j$, when it is $1$.

In order to simplify the notation, it is convenient to specify two additional matrices. The first one is 
the $R\times 1$ matrix given by
\begin{equation}{\label{pxlogm1}}
{\bf L}_{X} =\left[ 
\begin{array}{cccc}
-\log_2 P(x_1)   \\ 
-\log_2 P(x_2)   \\ 
\vdots    \\
-\log_2 P(x_R)   \\
\end{array}%
\right]. 
\end{equation}
The other one is the $S \times R $ matrix given by
\begin{equation}	{\label{pxylogm1}}
{\bf L}_{XY} =\left[
\begin{array}{cccc}
-\log_2 P(x_1/y_1) & -\log_2 P(x_2/y_1) & \cdots & -\log_2 P(x_R/y_1)  \\ 
-\log_2 P(x_1/y_2) & -\log_2 P(x_2/y_2) & \cdots & -\log_2 P(x_R/y_2)  \\ 
\vdots & \vdots & \ddots & \vdots  \\
-\log_2 P(x_1/y_S) & -\log_2 P(x_2/y_S) & \cdots & -\log_2 P(x_R/y_S)  \\
\end{array}
\right] .  
\end{equation}	
When $P(x_i)=0$ or $P(x_i/y_j)=0$ the corresponding entries in 
(\ref{pxlogm1}) or in (\ref{pxylogm1}) are zero. 

\subsection{The entropy block}
By considering the matrices defined in the previous section, in this 
block the goal is to obtain the entropies $H(X)$ and  $H(X/Y)$. 
	
The {\it a priori} entropy of the source $H(X)$ is a number (in bits units) 
calculated by the matrix multiplication,
\begin{equation}
H(X)= {\bf P}_X {\bf L}_{X}. 
\end{equation}			
					
The {\it a posteriori} entropy ${\bf H}_{X/Y} $ is an $ S \times 1$ 
matrix with entries given by
\begin{equation}
[{\bf H}_{X/Y}]_{j,1} = [{\bf P}^{T}_{X/Y}  {\bf L}_{XY}]_{j,j}, (j = 1, 2, ..., S),
\end{equation}
and the superscript $T$ stands for transpose. Since the output symbols 
$\left\{y_S\right\}$ occur with probabilities ${\bf P}_Y$, an average 
a-posteriori-entropy can be obtained by the following matrix multiplication
\begin{equation}
H(X/Y)={\bf P}_Y {\bf H}_{X/Y}. 
\end{equation}
This conditional entropy $H(X/Y)$ measures the average final uncertainty of 
$X$ after an observation of the output produced by the input. It is 
straightforward, by applying the proposed algorithm, to verify that these 
matrix multiplications presented here are equivalent to the mathematical 
expressions in~\cite{leebedford}.

\subsection{The information block} 
Following the flowchart in Fig.~\ref{fig.chart} the next step is to obtain the average 
information about $X$  conveyed through the channel by $Y$, namely $I(X,Y)$, 
which is given by $I(X,Y)=H(X)-H(X/Y)$, and the normalized symmetric uncertainty $U(X,Y)$. These quantitative parameters may be useful to compare accounting classifications in situations, such as, different accounting standards, periods, chart of account levels, and so on. 
\section{Application}{\label{sec.application}}
\begin{table}
\begin{center}
\caption{List of some economic events $x_i$, their frequencies ($P(x_i)$) and classifications ($y_i$).}
\begin{tabular}{ p{5.5cm} l| l l} 
  \hline                        
   &  & \multicolumn{2}{c}{Classification - $y_i$}\\ 
  $x_i$ & $P(x_i)$ & Debit & Credit \\  \hline
1- Unemployment Compensation&0.0014&01.01.01.001.00001&01.01.01.002.00005 \\
2- Sales in the State&0.0167&01.01.01.001.00001&03.01.01.001.00001 \\
3- Sales to other States&0.0153&01.01.01.001.00001&03.01.01.001.00002 \\
4- Resale of goods&0.0264&01.01.01.001.00001&03.01.01.001.00004 \\
5- Sales return&0.0014&01.01.01.001.00001&03.01.01.002.00007 \\
6- General Services sales&0.0125&01.01.01.001.00001&03.01.01.004.00001 \\
7- Loan Agreement&0.0014&01.01.01.002.00005&02.01.02.003.00001 \\
8- Buildings and Constructions&0.0014&01.03.02.001.00002&01.01.01.001.00001 \\
9- Equipment and machinery&0.0042&01.03.02.002.00005&01.01.01.001.00001 \\
10- Vehicles&0.0014&01.03.02.002.00007&01.01.01.001.00001 \\
11- Computing devices&0.0042&01.03.02.002.00008&01.01.01.001.00001 \\
$\vdots$ & $\vdots$ & $\vdots$ & $\vdots$ \\
78- 13th month´s salary&0.0014&04.01.03.004.00001&01.01.01.001.00001 \\
79- 13th month´s salary fees&0.0014&04.01.03.004.00001&02.01.03.002.00002 \\
80- 13th month´s salary taxes receivable&0.0014&04.01.03.004.00001&02.01.03.002.00004 \\
81- Water Costs &0.0042&04.01.03.004.00002&01.01.01.001.00001 \\
82- Accountant´s fee&0.0153&04.01.03.004.00005&02.01.03.001.00006 \\
83- Energy Costs&0.0056&04.01.03.004.00020&01.01.01.001.00001 \\
84- Office Supply&0.0042&04.01.03.004.00033&01.01.01.001.00001 \\
85- Telephony Costs&0.0097&04.01.03.004.00043&01.01.01.001.00001 \\
  \hline  
\end{tabular}
{\label{tab.events}}
\end{center}
\end{table}

In order to apply the theory to a realistic situation, we have investigated all economic events registered during the period of two consecutive years 
in a Brazilian small company located at S\~ao Paulo State. The data set contains 2075 daily transactions, 
with 1356 registered in the first year and 719 in the 
second year. The transactions and their accounting entries were coded according to 
a Brazilian specific chart of accounts with five levels. 

We obtain measures of information inherent to the structure of the company
chart of accounts. The company we are considering has a code with 
five pieces to discriminate uniquely their transactions. Thus, the classification process can be 
analyzed according to five different levels of aggregation. Table~\ref{tab.events} shows 
some economic events $x_i$, their relative frequencies $P(x_i)$, and their 
classification ($y_i$) according to the chart of accounts at Level 5. In the table, a typical classification $y_i$ is given by a pair of accounts, specified by the debt and credit columns. 

The probability of economic events $X$ that are classified as $Y$ is a matrix characterized 
by all entries being either one or zero, and having one, and only one, non-zero element 
in each row. At level 5 in the chart of accounts code each classification 
$y_i$ is associated to a single event. When the accounts are aggregated,
a single classification may be used to classify more than one event. However,
a specific event remains classified in only one $y_i$, and in this situation we have a deterministic accounting channel.
\begin{table}
\begin{center}
\caption{Conditional entropy $H(X/Y)$, mutual information $I(X,Y)$, and the symmetric uncertainty $U(X,Y)$ for 
the classification process according to the five levels contained in the Chart of Accounts (CA).}
\begin{tabular}{l | c c r |  c c r  } 
  \hline                        
  CA&	\multicolumn{3}{c|}{First year, $H(X) = 4.005$ bits} & \multicolumn{3}{c}{Second year, $H(X) = 
5.519$ bits} \\ \cline{2-7}
levels 	& $H(X/Y)$	& $I(X,Y)$ &	$U(X,Y)$ &  $H(X/Y)$	& $I(X,Y)$ &	$U(X,Y)$ \\ \hline
5 &  0.000 &	4.005 &	1.000 &	0.000	&	5.519	&	1.000 \\
4 &  	0.930 &	3.075 &	0.869 & 1.663	&	3.855	&	0.823 \\
3 &	1.460 &	2.545 &	0.777 & 2.406	&	3.112	&	0.721 \\
2 &	1.776 &  2.229 &	0.715 & 2.985	&	2.534	&	0.629 \\
1 &		1.827 &	2.178 &	0.705 & 3.066	&	2.453	&	0.611 \\
  \hline  
\end{tabular}
{\label{tab.results}}
\end{center}
\end{table}

\begin{figure}[!t] %F1
\centering\includegraphics[width=.8\textwidth] {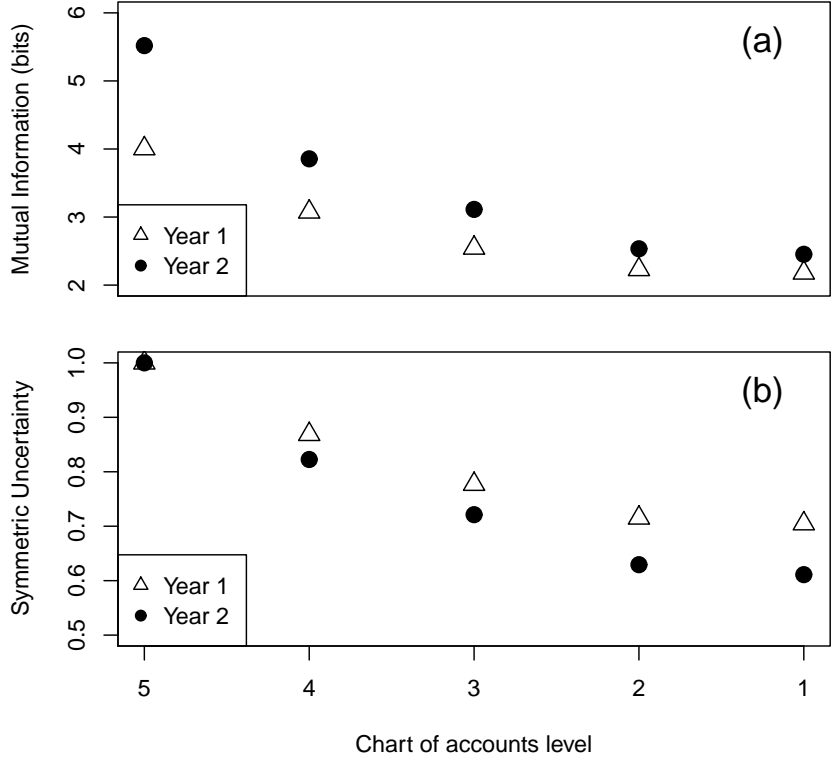}
\caption{Plot of the (a) mutual information, and (b) symmetric uncertainty, for the classification process at five levels according to the chart of accounts.}
\label{Fig.Ixy}\vspace*{-9pt}
\end{figure}
\vspace*{6pt}

To evaluate the information measures at each level of classification we use the algorithm developed in 
section~\ref{sec.algorith} carried out using the R-Soft\-ware (R-code is presented in the Appendix A). 
Table~\ref{tab.results} shows the values of $H (X )$, $H (X/Y )$, $I (X, Y )$, and $U (X, Y )$ at each 
level of classification for each year. The company has less variety of economic events in the first year 
($R_1 = 45$) compared to the second one ($R_2= 85$), resulting in a lower entropy $H (X)$ for the first 
year. An increase of $H (X/Y )$ from Level 5 to Level 1 is observed since uncertainty about the economic 
events increases as the classifications are performed at more aggregated levels. The loss of information 
relative to the economic events is given by $I(X,Y )$ and $U(X,Y)$, and whose decreasing values from 
Level 5 to Level 1 indicate a reduction of association between economic events and their classifications 
at more aggregated levels. For a specific year, the value of $I (X, Y)$ provides an information criterion 
to compare accounting classifications at different levels in the chart of accounts. To compare different 
time periods it is better to consider the normalized metric $U (X, Y )$, which can be used as an index 
indicating the strength of dependence between the economic events and their classifications. The Level 5 
has $U (X, Y ) = 1$ since each classification corresponds to a specific economic event. Then, $U(X,Y)$ 
decreases relatively to Level 5,  attaining its minimum value at Level 1, which has the smaller number of classification states. Compared at Level 1, the symmetric uncertainty for the first and the second years 
is $U (X, Y) = 0.705$ and $U (X, Y) = 0.611$, respectively. Therefore, for this company, the first year has 
a higher degree of dependence between economic events and its classifications. In Figure~\ref{Fig.Ixy}a 
and \ref{Fig.Ixy}b, values of $I(X,Y)$ and $U(X,Y)$ were plotted for each level of classification, respectively. Differences over the years are expected once the company changes its set of activities. 
The more aggregated the classification, the closer the values of mutual information. However, the study 
of more aggregate levels is useful for improving financial statement analysis. From Fig.~\ref{Fig.Ixy}b we 
see that differences over periods at more aggregated levels are evidenced by the values of $U(X,Y)$. In 
particular, the first level of classification in the chart of accounts should contain only the main 
accounts, such as assets, liability, equity, and so on, which is common to all accounting standards. 
Then the symmetric uncertainty at the first level of classification can be used as an index in a financial 
statement analysis to distinguish between classifications over the years. We also suggest $U (X, Y)$ as 
an index to compare the information in the accounting classification for different sectors, or different 
accounting choices and standards.

\section{Concluding remarks}{\label{sec.conclusion}}

In this work a matrix formalism for the information theory formulation of the 
accounting classification process was presented. An 
algorithm generalizing the procedure of Lee and Bedford~\cite{leebedford} 
for the calculation of information theoretic measures in the accounting classification process was proposed. The algorithm provided a 
matrix procedure suitable for software implementation integrated to 
information systems.  The algorithm was applied to evaluate
information theoretic measures, mutual information and symmetric uncertainty, for daily transactions recorded 
by a small company during two years. We have verified an information loss 
inherent to aggregation of levels in the chart of accounts. To the best of 
our knowledge, this is the first calculation of information theoretic measures 
for the accounting classification process of a company. In particular, the symmetric uncertainty at the first level of
classification in a chart of accounts seems to be a useful parameter
for comparing companies over time or in different sectors or different accounting choices and standards. Furthermore, the accounts at the chart of accounts first level are commonly used to form financial and economic indexes to characterize the firm. Since the symmetric uncertainty contains the proportion of information shared between the main accounts and all the economic events, it can be used itself as a global index associated to the company. The relation between symmetric uncertainty and the financial and economic indexes deserves a further study and will be addressed in a future work.

It is worthwhile to mention that the probabilities used by the proposed 
procedure were entirely based on observed frequencies. On the other hand, 
the probabilities can be considered as parameters in a theoretical analysis 
of different classification standards and may be a useful quantitative tool 
in the searching for the {\it a priori} most adequate level of classification.

We hope that this work may contribute to highlight Accounting as an interesting field for interdisciplinary research, and also to renew and stimulate the application of information theoretic tools in the accountancy practice.

\vspace{1.0cm}

\appendix{\bf{Appendix A. Algorithm in R language}}

In the following we transcribe the $R$ code implemented to obtain the 
mutual information. Input files can be obtained by e-mail to the authors.

\begin{verbatim}
# Algorithm for mutual information calculation 
#
# The inputs are in the csv file -----
freqs <- read.table("inputs.csv",header=TRUE,sep=";")
attach(freqs)
Tot <- length(freqs)
Tot
# Tot are the number of columns in the dataframe "freqs" 
#
# The Probability Block --------------
#
# Px: the probability of economic events, 
# the first dataframe column:
R <- length(Freqx)
R
Px <- matrix(Freqx,nrow=1,ncol=R,byrow=TRUE)
#
# PyGx Probability of y given x, 
# The last S dataframe columns:
S = Tot - 1
S
PyGx <- data.matrix(freqs[,2:Tot])
#
# Py: probability of classification y
Py <- Px %*% PyGx
#
# Additional Matrices: Pxd e Pydm1
Pxd <- diag(Freqx)    
Freqy <- colSums(Py)
Freqy2 <- 1/Freqy
Pydm1 <- diag(Freqy2) 
#
# PxGy: probability of x, given y
Mult1 <- PyGx %*% Pydm1
PxGy <- Pxd %*% Mult1
#
# Additional matrix: Lx
FreqxLog2m <- -log2(Freqx)
Lx <- matrix(FreqxLog2m,nrow=R,ncol=1)
#
# Additional matrix: Ly
FreqyLog2m <- -log2(Freqy)
Ly <- matrix(FreqyLog2m,nrow=S,ncol=1)
#
# Transpost of PxGy
PxGyT <- t(PxGy)
#
# Additional matrix: Lxy
Lxy <- PxGy
for (i in 1:R){
for (j in 1:S){
if (Lxy[i,j] != 0){Lxy[i,j] <- -log2(Lxy[i,j]) }
}
}
#
# The Entropy Block ------------------
# H(x): Entropy for the economic events
Hx <- Px %*% Lx
print(Hx,digits=12)
#
# H(y): Entropy for classifications (not used)
Hy <- Py %*% Ly
print(Hy,digits=12)
#
# Conditional entropy (vector), HxGy
HxGy <- diag(PxGyT %*% Lxy)
#
# Conditional entropy (average), H(X/Y):
HxBy <- Py %*% HxGy
print(HxBy,digits=12)
#
# The Information Block --------------
# Ixy: the mutual information
Ixy <- Hx - HxBy
print(Ixy,digits=12)
#
# Uxy: Symmetric uncertainty
Uxy = 2.0*Ixy/(Hx+Hy)
print(Uxy,digits=12)
#
detach(freqs)
rm(freqs,i,j,R,S,Tot,Hx,Hy,HxGy,HxBy,Ixy,Uxy,Mult1)
rm(Px,Pxd,PxGy,Lxy,PxGyT,Lx,Ly,Py,Pydm1,PyGx)
rm(FreqxLog2m,Freqy,Freqy2,FreqyLog2m)
ls()
\end{verbatim}


\section*{References}
\bibliographystyle{elsarticle-num} 
\bibliography{references}

\end{document}